\documentclass[aps,prd,email,preprint,showpacs,showkeys,preprintnumbers,amsmath,amssymb,nofootinbib]{revtex4}
\usepackage{color}
\usepackage{latexsym}
\usepackage{amsmath}
\usepackage{amssymb}
\usepackage{eufrak}
\usepackage{euscript}
\usepackage{pstricks}
\usepackage{graphics}
\usepackage{graphicx}
\usepackage{picture}
\usepackage{pstricks,pst-coil}
\usepackage{pst-all}

\newcommand{\be}{\begin{equation}}
\newcommand{\ee}{\end{equation}}
\newcommand{\bn}{\begin{eqnarray}}
\newcommand{\en}{\end{eqnarray}}

\def\[{\left\lbrack}
\def\]{\right\rbrack}

\def\({\left(}
\def\){\right)}
\def\ni{\noindent}	
\def\MyItem[#1]#2{\item[{#1}]#2}

\def\[{\left\lbrack}
\def\]{\right\rbrack}

\def\({\left(}
\def\){\right)}

\def\uma{\rm 1\!\!\hskip 1 pt l}
\begin{document}


\title{\Large{Quantum electrodynamics and the electron self-energy in a deformed space with a minimal length scale}}

\author{Apollo V. Silva$^{a}$}
\email{apollosilva@ufrrj.br}
\author{E. M. C. Abreu$^{b,c}$}
\email{evertonabreu@ufrrj.br}
\author{M. J. Neves$^{b}$}
\email{mariojr@ufrrj.br}

\affiliation{$^a$Departamento de F\'{i}sica, Universidade Federal Rural do Rio de Janeiro, BR 465-07, 23890-971, Serop\'edica, RJ, Brazil\\
$^b$Grupo de F\' isica Te\'orica e Matem\'atica F\' isica, Departamento de F\'{\i}sica, \\
Universidade Federal Rural do Rio de Janeiro\\
BR 465-07, 23890-971, Serop\'edica, RJ, Brazil \\\\
$^c$Departamento de F\'{\i}sica, ICE, Universidade Federal de Juiz de Fora,\\
36036-330, Juiz de Fora, MG, Brazil\\\\
\today\\}

\begin{abstract}
\ni The main motivation to study models in the presence of a minimal length is to obtain a quantum field theory free of the divergences.  In this way, in this paper, we have constructed a new framework for quantum electrodynamics embedded in a minimal length scale background. New operators are introduced and the Green function method was used for the solution of the field equations, i.e., the Maxwell, Klein-Gordon and Dirac equations.   We have analyzed specifically the scalar field and its one loop propagator.  The mass of the scalar field regularized by the minimal length was obtained.   The QED Lagrangian containing a minimal length was also constructed and the divergences were analyzed.
The electron and photon propagators, and the electron self-energy at one loop as a  function of the minimal length was also obtained.
\end{abstract}


\maketitle


\section{Introduction}

Several studies in string theory \cite{string}, noncommutative geometry \cite{cls} and loop quantum gravity \cite{garay} suggest the existence of a minimal length.   The main argument would be that at the Planck scale, the Heisenberg uncertainty principle would be substituted by the generalized uncertainty principle \cite{string}.

The necessity of introducing a minimal length scale in physical theories is an old concept since the thirties \cite{Heisenberg38}.  In 1938, Heisenberg have published this idea as a way to cure the divergence problems that have been dwelled in QFT computation since the earlier times.  For example, in the scattering process computations in QED, or, as another example, in order to calculate the Compton effect scattering cross section, the final result showed infinite values, which is a nightmare if you want to obtain a numerical result.   In this scenario, the minimal length scale would be a universal constant.  As well known, the vacuum velocity of light ($c$) represents the scale of velocity where the effects of the special relativity are relevant, as well as the Planck constant ($\hbar$) represents the microscopic scale governing the quantum world, where the effects of quantum mechanics are important.  Another well known constant is the gravitational constant ($G$), which measures the gravitational interactions between masses.

In 1900, Planck has suggested that the combination between the underlying constants of Nature, i.e., $c$, $\hbar$ and $G$, would form the length scale
\begin{eqnarray}\label{1}
\ell_{P}=\sqrt{\frac{\hbar G}{c^3}} \simeq 10^{-33} \, \mbox{m} \; ,
\end{eqnarray}
where $\ell_p$ is the Planck length. Eq. (\ref{1}) shows that the unification of the three underlying constants demonstrates that the Planck length must be the proper scale where the gravitational effects in the quantum world are relevant.

The first physical theory in which it was introduced a length scale was based in the notion of the noncommutativity of space-time, where the coordinates $x^{\mu}\, (\mu=0,1,2,3)$ are promoted to hermitian operators in quantum mechanics, i.e., $\hat{x}^{\mu}$, do not commute, and have the following commutation relation
\begin{eqnarray} \label{2}
\left[\hat{X}^{\mu},\hat{X}^{\nu} \right]=i \, \ell^2 \, \theta^{\mu\nu} \, \hat{\uma} \; ,
\end{eqnarray}
\ni where $\ell$ is a length scale, and $\theta^{\mu\nu}$ is a constant $4\, \times \, 4$ antisymmetric matrix.  Notice that at the limit where $\ell \, \rightarrow \,0$, the standard quantum theory can be obtained.  In this scenario the space-time coordinates are observables that do not commute.  This fact changes the understanding of the space-time geometry in this length scale. As a consequence, the concept of position measure in this space-time would change also, where the Heisenberg uncertainty principle must be satisfied
\begin{eqnarray} \label{3}
\Delta \hat{X}^{\mu} \Delta \hat{X^{\nu}} \gtrsim \ell^{2} \, \theta^{\mu\nu} \; .
\end{eqnarray}

\ni However, a short time later, C. N. Yang has shown that Snyder's theory did not solve the infinities problems in QFT.  This fact has led the NC  theory to a deep sleep for about fifty years. Only after an important result that has came from string theory \cite{seibergwitten99}, where a string embedded into a magnetic background, shows NC features concerning the space coordinates. This result made the noncommutativity issue to be a top subject in theoretical physics \cite{nekrasov,szabo,abreuneves} during the last years.

In current days, new analysis of string theory and quantum gravity, suggest us the existence of a minimal length in Nature \cite{Szabo03}.  This conclusion has motivated the reformulation of quantum mechanics together with the gravitation theories and QFT, which were built only in the presence of a minimal length scale.

Another possibility of introducing a minimal length scale $\ell$ leaves the direct modification of the uncertainty relation in quantum mechanics \cite{Moyaedi2013}, namely,
\begin{eqnarray} \label{RelIncl}
\Delta \hat{X} \Delta \hat{P} \geq \frac{\hbar}{2} \left[1+ \frac{\ell^{2}}{\hbar^{2}} \, (\Delta \hat{P})^{2} \right] \; ,
\end{eqnarray}
\ni where the particle position and momentum must be changed within a quantum theory scenario where the scale $\ell$ must be relevant.  Eq. (\ref{RelIncl}) is the well known generalized uncertainty principle, which can be seen also (as we have just said) as the origin of the appearance of a minimal length since it can be written in an even more general way such as

\begin{eqnarray} \label{RelIncl2}
\Delta \hat{X} \Delta \hat{P} \geq \frac{\hbar}{2} \left[1+ \frac{\ell^{2}}{\hbar^{2}} \, (\Delta \hat{P})^{2} \,+\, \gamma\,\, \right] \; ,
\end{eqnarray}

\ni where $\ell$ and $\gamma$ are positive and independent parameters of $\Delta \hat{X}$ and $\Delta \hat{P}$.  
However, they can, in general, rely on the expectation values $\langle \hat{X} \rangle$ and $\langle \hat{P} \rangle$. 
The standard Heisenberg uncertainty relation can be
recovered from (\ref{RelIncl2}) by doing $\ell\, =\, \gamma\, =\, 0$. It is direct to realize that (\ref{RelIncl2}) means a minimum
position uncertainty of $\left(\Delta \hat{X}\right)_{min} = \ell$.

In fact, the modified uncertainty relation (\ref{RelIncl}) is the beginning of an effective model and it must have corrections at higher orders like
the one in the following series
\begin{eqnarray}\label{RelInclserie}
\Delta \hat{X} \Delta \hat{P} \geq \frac{\hbar}{2} \left[1+ c_1 \, \frac{\ell^{2}}{\hbar^{2}} \, \left(\Delta \hat{P}\right)^{2}
+c_{2} \, \frac{\ell^{4}}{\hbar^{4}} \, \left(\Delta \hat{P}\right)^{4}+ \ldots \right]=\frac{\hbar}{2} \sum_{n=0}^{\infty} c_{n} \left( \frac{\ell^{2}}{\hbar^{2}}
\, \Delta\hat{P}^2 \right)^{n} ,
\end{eqnarray}
where $c_{n}$ are real constants. Since we have very small terms, we propose the series above becomes a general uncertainty relation given
by the exponential function
\begin{eqnarray}\label{RelInclexp}
\Delta \hat{X} \Delta \hat{P} \geq \frac{\hbar}{2} \; \exp \left[ \frac{\ell^{2}}{\hbar^{2}} \, \left(\Delta \hat{P}\right)^{2} \right] \; .
\end{eqnarray}
In this scenario, it is easy to see that the modified uncertainty relation lead us to a measure of the minimum uncertainty for the position given by
\begin{eqnarray} \label{5}
\left(\Delta \hat{X} \right)_{min} \! = \sqrt{e/2} \; \ell \simeq \ell \; ,
\end{eqnarray}
\ni where $\ell$ acts as the minimal length scale, and any measure of the system can not be less than its value, which corresponds to the maximum
value for the linear momentum given by
\begin{eqnarray} \label{6}
\left(\Delta \hat{P} \right)_{max}= \frac{\hbar}{\sqrt{2} \, \ell} \; .
\end{eqnarray}
\ni Hence, the maximum energy scale relative to these values is
\begin{eqnarray} \label{7}
\left(\Delta E\right)_{max}=\frac{\hbar c}{\sqrt{2} \, \ell} \; .
\end{eqnarray}
\ni Since $\ell$ has a very small value, the minimum uncertainty for the value of the energy has a very high value, if it can be compared to the energy scale of the modern particle accelerators.

The organization of this paper follows the sequence such that in section $2$ we show both the algebra and representation of the operators.
In section $3$, we have reviewed the electromagnetism with a minimal length. In section $4$, we have analyzed the electrostatic properties in the presence of the minimal length scale. In section $5$, the scalar model was introduced, and the influence of $\ell$ on field mass was discussed.
In section $6$ we have proposed the construction of a new effective quantum electrodynamics. The conclusions were depicted in section $7$.

\pagebreak

\section{The algebraic structure and observable representation}

A few years back, Quesne and Tkachuk had constructed a Lorentz deformed algebra which is covariant and described a quantized space-time with $D\,+\,1$ dimensions.

The uncertainty relation (\ref{RelIncl}) implies that the observables of position $\hat{X}^{\mu}$, and momenta
$\hat{P}^{\mu}$, must satisfy to algebraic structure of this Quesne-Tkachuk algebra \cite{Quesne2006JPA,Quesne2006Csz}.

\begin{eqnarray}\label{Relcomutabeta}
\left[\hat{X}^{\mu},\hat{X}^{\nu}\right]\! &=& \! i\hbar \, \frac{2\beta-\beta^{\prime}-\left(2\beta+\beta^{\prime} \right) \beta\hat{P}^{2} }{1-\beta\,\hat{P}^{2}} \, \left( \hat{P}^{\mu} \, \hat{X}^{\nu} - \hat{P}^{\nu} \, \hat{X}^{\mu} \right) \; ,
\nonumber \\
\left[\hat{X}^{\mu},\hat{P}^{\nu}\right] \! &=& \! i\hbar \left[\left(1-\beta \, \hat{P}^{2} \right) \, \eta^{\mu\nu}
- \beta^{\prime} \, \hat{P}^{\mu} \, \hat{P}^{\nu}   \right] \; ,
\nonumber \\
\left[\hat{P}^{\mu},\hat{P}^{\nu}\right] \! &=& \! 0 \; ,
\end{eqnarray}
where $\beta$ and $\beta^{\prime}$ are real non-negative parameters ($\beta ,\,\beta'\,\geq 0$), that have dimension of momentum to $-2$, i.e., (momentum)$^{-2}$, and they are related to minimal length scale and Planck constant. For simplicity, we have defined $\hat{P}^{2}:=\hat{P}_{\rho}\hat{P}^{\rho}$.
When $\beta=\beta^{\prime}=0$, it is direct that this algebra reduces to the usual commutation relations
\begin{eqnarray}\label{RelcomutaMQ}
\left[\hat{X}^{\mu},\hat{P}^{\nu}\right]= i\hbar \, \eta^{\mu\nu}
\hspace{0.3cm} , \hspace{0.3cm}
\left[\hat{X}^{\mu},\hat{X}^{\nu}\right]=0
\hspace{0.3cm} , \hspace{0.3cm}
\left[\hat{P}^{\mu},\hat{P}^{\nu}\right]=0 \; .
\end{eqnarray}

The first commutation relation in (\ref{Relcomutabeta}) is an approach of a NC space-time. For special case,
$\beta=0$, and $\beta^{\prime} \ll 1$, the Snyder's NC space-time is obtained
\begin{eqnarray}\label{RelcomutaSnyder}
\left[\hat{X}^{\mu},\hat{X}^{\nu}\right]\! &=& \! -i\hbar \, \beta^{\prime} \left( \hat{P}^{\mu} \, \hat{X}^{\nu} - \hat{P}^{\nu} \, \hat{X}^{\mu} \right) \; ,
\nonumber \\
\left[\hat{X}^{\mu},\hat{P}^{\nu}\right]\! &=& \! i\hbar \, \left( \, \eta^{\mu\nu}
- \beta^{\prime} \, \hat{P}^{\mu} \, \hat{P}^{\nu}   \right) \; ,
\nonumber \\
\left[\hat{P}^{\mu},\hat{P}^{\nu}\right]\! &=& \! 0 \; .
\end{eqnarray}
Therefore, the altered uncertainty relation implies in a NC space-time structure.
Here we are interested in studying the model using the commutative space-time approach.
For this goal to be achieved, we have to choose the condition $\beta^{\prime}=2\beta$, where $\beta:=\ell^{2}/\hbar^{2}$.
Thus, the previous algebra will be given by
\begin{eqnarray}\label{Relcomutal1}
\left[\hat{X}^{\mu},\hat{P}^{\nu}\right] \! &=& \! i\hbar \left[\left(1- \frac{\ell^{2}}{\hbar^{2}} \, \hat{P}^{2} \right) \, \eta^{\mu\nu}
- \frac{2\ell^{2}}{\hbar^{2}} \, \hat{P}^{\mu} \, \hat{P}^{\nu}   \right]   \; ,
\nonumber \\
\left[\hat{X}^{\mu},\hat{X}^{\nu}\right] \! &=& \! 0
\hspace{0.3cm} , \hspace{0.3cm}
\left[\hat{P}^{\mu},\hat{P}^{\nu}\right]=0 \; .
\end{eqnarray}
Consequently, we obtain the minimal position uncertainty
\begin{eqnarray}\label{IncX}
\left(\Delta X^{i}\right)_{min}=\sqrt{5} \, \ell \, \left[1-\frac{\ell^2}{\hbar^2} \, \langle \left( \hat{P}^0 \right)^2 \rangle \right]^{1/2}
\simeq \sqrt{5} \, \ell
\hspace{0.2cm} , \hspace{0.2cm}
i=1,2,3 \; .
\end{eqnarray}
The algebra (\ref{Relcomutal1}) suggests us to introduce the representation for the  momentum operator as being
\begin{eqnarray}\label{Pmul2}
\hat{P}^{\mu} \longmapsto \left(1+ \ell^{2} \, \Box\right) \, i \hbar \, \partial^{\mu}  \; ,
\end{eqnarray}
where the position operator remains unaltered, that is, $\hat{X}^{\mu} \longmapsto \hat{x}^{\mu}$, and the usual relation is
$[\hat{x}^{\mu},i \hbar \, \partial^{\nu}]=i \hbar \, \eta^{\mu\nu} \, \hat{\uma}$. The quantum mechanics based on
previous representation is well established in the literature, see \cite{Kempf94,Kempf95,Kempf96}.
The harmonic oscillator is invariant under translational symmetry, see \cite{Kempf97,Chang2002}.

Now we will propose to study a representation of momentum operator for the general case where we will introduce the exponential operator
\begin{eqnarray}\label{Pmunablamu}
\hat{P}^{\mu} \longmapsto e^{ \ell^{2} \, \Box} \, i \hbar \, \partial^{\mu}  \; ,
\end{eqnarray}
where, if we consider $\ell^2 \ll 1$, we have that $$e^{ \ell^{2} \, \Box} \simeq 1 + \ell^{2} \, \Box\,\,,$$ and, in Thai way, the operator (\ref{Pmunablamu})
is reduced to (\ref{Pmul2}). As consequence, the commutation relation involving $\hat{X}^{\mu}$, and $\hat{P}^{\mu}$ is given by
\begin{eqnarray}\label{Relcomutal}
\left[\hat{X}^{\mu},\hat{P}^{\nu}\right] = i\hbar \, \eta^{\mu\nu} \, e^{\frac{1}{2} \, W\left(-\frac{2\ell^{2}}{\hbar^2} \, \hat{P}^2\right)} \, \hat{\uma}
- \frac{i2\ell^{2}}{\hbar} \, \hat{P}^{\mu} \, \hat{P}^{\nu} \, e^{-\frac{1}{2} \, W\left(-\frac{2\ell^{2}}{\hbar^2} \, \hat{P}^2\right)}  \; ,
\end{eqnarray}
where $W$ is known as the Lambert function, or Product Logarithm. It is clear that for $\ell^2 \ll 1$, the algebra (\ref{Relcomutal1}) is recovered.   %
%
%
For convenience, we write the momentum operator as $\hat{P}^{\mu} \longmapsto i \hbar \, \nabla^{\mu}$,
where $\nabla^{\mu}:=e^{ \ell^{2} \, \Box} \, \partial^{\mu}$.
%
%

Our motivation here is to stress that the values of position and momentum, in a quantum length scale, must rely on a new length scale, which candidate is the Planck length, i.e., $\ell_{P} \sim 10^{-33}\, cm$.  In this scale, the current QFT is not sufficient in order to describe the fundamental interactions that live in Nature.   So, we would need a new QFT.   In this scenario, the length scale can be introduced directly in the uncertainty relation, and under some conditions, it can lead us to a Lee-Wick \cite{LeeWick69} kind of QFT, where the mass of the Lee-Wick field depends on the length scale. Due to this length scale, the first modification happen in quantum mechanics, where the momentum operator can have the form
\begin{eqnarray} \label{8}
\hat{P}_{\mu} \longmapsto i \hbar \nabla_{\mu} \; ,
\end{eqnarray}
\ni where $\nabla_{\mu}$ is defined by
\begin{eqnarray} \label{9}
\nabla_{\mu}= e^{\ell^{2} \, \Box} \, \partial_{\mu} \; ,
\end{eqnarray}
\ni where $\ell$ is the  same parameter with the length dimension mentioned above. In the limit $\ell \rightarrow 0$, we can obtain the standard quantum mechanics operator
\begin{eqnarray}\label{10}
\hat{P}_{\mu} \longmapsto i \hbar \partial_{\mu} \; .
\end{eqnarray}
\ni Hence, the construction of the QFT models that will be carried out from now on, will be based on the introduction of a new momentum operator in the field equation, such as the Klein-Gordon, Dirac and the Maxwell equations. In the next section, we start the discussion concerning the field equations and the classical electromagnetism equations.


%
%
\section{The electromagnetism in the presence of a minimal length}
Let us now make a classic analysis concerning the electromagnetism in the presence of a length scale, where the main problem is to realize how the charged particles interact. Our task is to fathom such interaction, following the basic Maxwell equations (ME), which can be modified by the introduction of a length scale $\ell$. First of all, it is important to mention that the Lagrangian that provides the field equations, following the minimum action, is given by
\begin{eqnarray} \label{LEMnablaF}
{\cal L}_{EM}=-\frac{1}{4 \mu_{0}} \, F_{\mu\nu}F^{\mu\nu}-J_{\mu}A^{\mu} \; ,
\end{eqnarray}
\ni where $\mu ,\nu = \{ 0,1,2,3 \}$ are the space-time indices and $F_{\mu\nu}=\nabla_{\mu} A_{\nu}-\nabla_{\nu} A_{\mu}$ is the electromagnetic field tensor, which is an antisymmetric second order tensor, which components are the electric and magnetic fields, and $J^{\mu}=(\rho,\vec{J})$ represents the  model charges and current distribution.

Explicitly, the Lagrangian (\ref{LEMnablaF}) can be written as :
\begin{eqnarray}\label{LEMnabla}
{\cal L}_{EM}=-\frac{1}{4 \mu_{0}} \, e^{\ell^{2}\Box}f_{\mu\nu}e^{\ell^{2}\Box}f^{\mu\nu} -J_{\mu}A^{\mu} \; ,
\end{eqnarray}
where $f_{\mu\nu}:=\partial_{\mu}A_{\nu}-\partial_{\nu}A_{\mu}$. Using the approximation, $e^{\ell^{2}\Box}\simeq 1+\ell^2\Box$, we can obtain
the Lagrangian up to order $\ell^2$, which is given by
\begin{eqnarray}
{\cal L}_{EM} \! &\simeq& \! -\frac{1}{4 \mu_{0}} \, \left(1+\ell^2\Box\right)f_{\mu\nu}
\left(1+\ell^2\Box\right)f^{\mu\nu} -J_{\mu}A^{\mu}
\nonumber \\
\! & \simeq & \! -\frac{1}{4 \mu_{0}} \, f_{\mu\nu}\left(1+2\ell^2\Box \right)f^{\mu\nu}-J_{\mu}A^{\mu} \; .
\end{eqnarray}
where $f^{\mu\nu}:=\partial^{\mu}A^{\nu}-\partial^{\nu}A^{\mu}$. This is like Lee-Wick's electrodynamics, in which
the Lee-Wick mass is determined by minimal length scale according to the relation
\begin{eqnarray}
m_{LW}=\frac{\hbar}{\sqrt{2} \, \ell c} \; .
\end{eqnarray}
A recent estimative of the Lee-Wick mass is $m_{LW} \simeq 410 \; \mbox{GeV}/c^2$ \cite{NevesTurcati2014}, which implies that
$\ell < 3.4 \times 10^{-20} \, \mbox{m}$. For some application of Lee-Wick model, see \cite{Accioly2011}.
The variational principle leads us to the equation
\begin{eqnarray}\label{12}
\nabla_{\mu}F^{\mu\nu}=\mu_{0} \, J^{\nu} \; ,
\end{eqnarray}
\ni and the field tensor satisfies the Bianchi equation
\begin{eqnarray}\label{13}
\nabla_{\mu}F_{\nu\rho}+\nabla_{\nu}F_{\rho\mu}+\nabla_{\rho}F_{\mu\nu}=0 \; .
\end{eqnarray}
\ni If we substitute the definition  of the field tensor, the equation for the field $A^{\mu}$ is given by
\begin{eqnarray} \label{14}
\left(\nabla_{\mu}\nabla^{\mu}\right) A^{\nu}(x)- \nabla^{\nu}\left(\nabla_{\mu}A^{\mu}\right)=\mu_{0} \, J^{\nu}(x) \; .
\end{eqnarray}
The gauge invariant models, namely, the tensor $F^{\mu\nu}$ is invariant under the gauge transformation
\begin{eqnarray} \label{15}
A'^{\mu}=A^{\mu}-\nabla^{\mu}\lambda(x) \; ,
\end{eqnarray}
so, if we fix the Lorenz gauge, $\nabla_{\mu} A^{\mu}=0$, the field equation for $A^{\mu}$ is
\begin{eqnarray}\label{16}
e^{2\ell^{2}\, \Box} \; \Box A^{\mu}(x)=\mu_{0} \, J^{\mu}(x) \; .
\end{eqnarray}
Therefore, we will analyze the static case, as a particular case of (\ref{16}). We are interested in the understanding of the
static interaction between charged particles, and to compare it to the Coulomb case. This behavior can reveal us the inner structure of charged particles.

\section{The electrostatic in the presence of a length scale}

Now we will simplify the difficult problem mentioned above.  The electrostatic case is the simpler way to understand how the electromagnetic interaction is affected thanks to a new minimum length scale.  Concerning the static case, the potential equations (\ref{16}) lead us to the Poisson equation for the scalar potential
\begin{eqnarray}\label{17}
e^{-2\ell^{2}\vec{\nabla}^{2}} \vec{\nabla}^{2}\Phi(\vec{r}) = -\frac{1}{\epsilon_0} \, \rho(\vec{r}) \; .
\end{eqnarray}
\ni This, of course, is a partial differential equation, non-homogeneous and of infinite order which involves an exponential expansion following a Taylor series
\begin{eqnarray}\label{18}
\left(1-2\ell^{2}\vec{\nabla}^{2}+\frac{\left(2\ell^{2}\right)^{2}}{2!} \, \vec{\nabla}^{4} + \ldots  \right) \nabla^{2}\Phi( \vec{r} ) = -\frac{1}{\epsilon_0} \, \rho( \vec{r} ) \; .
\end{eqnarray}
\ni To solve this equation we can use the Green function method.  Hence, we can write the solution as
\begin{eqnarray}\label{19}
\Phi(\vec{r})=\int_{{\cal R}} d^3 \vec{r}^{ \; \prime} \, G(\vec{r} - \vec{r} \, ') \left( -\frac{\rho(\vec{r}\,')}{\epsilon_0} \right) \; ,
\end{eqnarray}
\ni where $G (\vec{r}-\vec{r}\,')$ is the Green function of our problem. In order to consider the integral in Eq. (\ref{19}) as a solution of the Poisson equation, the Green function has to satisfy the following equation
\begin{eqnarray}\label{20}
e^{- {2{\ell}^2 {\nabla}^2}} {\nabla}^2 G(\vec{r} - \vec{r}\,') = \delta^3(\vec{r} - \vec{r}\,') \; ,
\end{eqnarray}
\ni where $\delta^3 (\vec{r}-\vec{r}\,')$ is the Dirac delta function in $D=3$.  Solving Eq. (\ref{20}) by using the Fourier transform method we have that it
%
%
%
%
leads us to the integral form such as
%
%
%
%
%
\begin{eqnarray}\label{25}
G(\vec{r} - \vec{r}\,') = \int \frac{d^3 \vec{k}}{(2\pi)^3} \; e^{- i \vec{k} \cdot (\vec{r} - \vec{r}\,')} \left( - \frac{e^{-2{\ell}^2 {\vec{k}}^2}}{{\vec{k}}^2 } \right) \; .
\end{eqnarray}
Using spherical coordinates, the integral in (\ref{25}) can be written as
\begin{eqnarray} \label{26}
G(\vec{r} - \vec{r}\,') = - \frac{1}{2 {\pi}^2 \, |\vec{r} - \vec{r}\,'|} \, \int_0^\infty dk \; e^{-2{\ell}^2 k^2} \, k^{-1} \, \sin\left(k |\vec{r} - \vec{r}\,'|\right)  \; .
\end{eqnarray}
\ni The radial integral in $k$ can be written \cite{Gradshteyn} as an error function (Erf)
\begin{eqnarray}\label{30}
G(\vec{r}-\vec{r}^{\; \prime})=- \frac{1}{4 \pi |\vec{r}-\vec{r}^{\; \prime}|} \; \mbox{Erf}\left(\frac{|\vec{r}-\vec{r}^{\; \prime}|}{2\sqrt{2} \, \ell} \right) \; .
\end{eqnarray}
\ni Consequently, the solution considering the electrostatic potential can be written as
\begin{eqnarray}\label{31}
\Phi(\vec{r}) = \frac{1}{4\pi \epsilon_{0}} \int_{{\cal R}} d^3 \vec{r}\;' \; \frac{\rho(\vec{r}\,')}{|\vec{r}-\vec{r}\,'|} \; \mbox{Erf}\left(\frac{|\vec{r}-\vec{r}\,'|}{2\sqrt{2} \, \ell} \right) \; .
\end{eqnarray}
\ni and the electrostatic field is given by
\begin{eqnarray}\label{32}
\vec{E}(\vec{r})= \frac{1}{4\pi \epsilon_{0}} \int_{{\cal R}} d^3 \vec{r}\,' \; \frac{\rho(\vec{r}\,')}{|\vec{r}-\vec{r}\,'|^2} \; \frac{\vec{r}-\vec{r}\,'}{|\vec{r}-\vec{r}\,'|} \; \mbox{Erf}\left(\frac{|\vec{r}-\vec{r}\,'|}{2\sqrt{2} \, \ell} \right)
\nonumber \\
 -\frac{1}{4\pi \epsilon_{0}} \, \frac{1}{\ell\sqrt{2\pi}} \int_{{\cal R}} d^3 \vec{r}\,' \; \frac{\rho(\vec{r}\,')}{|\vec{r}-\vec{r}\,'|} \; \frac{\vec{r}-\vec{r}\,'}{|\vec{r}-\vec{r}\,'|} \, e^{-\left(\frac{|\vec{r}-\vec{r}\,'|}{ 2\sqrt{2} \,\ell}\right)^{2}}  \; .
\end{eqnarray}
%
%
%
%
%
\ni In the limit $\ell \rightarrow 0 $, considering the error function properties, it is easy to see that
\begin{eqnarray}\label{35}
\lim_{\ell \rightarrow 0} \, \mbox{Erf}\left(\frac{|\vec{r}-\vec{r}\,'|}{2\sqrt{2} \, \ell}\right) = 1 \; ,
\end{eqnarray}
\ni and the expressions for the potential and for the electrostatic case turn to the standard case.

Considering the point charge, we can use the charge distribution at the origin, $\rho \left(\vec{r}\,'\right) = q \, \delta^3 \left(\vec{r}\,'\right)$,
in order to obtain the potential given by
\begin{eqnarray}\label{36}
\Phi(r) = \frac{q}{4 \pi \epsilon_{0} \, r} \; \mbox{Erf}\left(\frac{r}{2\sqrt{2} \, \ell} \right) \; ,
\end{eqnarray}
\ni where $r$ is the distance between the charge and a generic point in the space.
\begin{figure}[h]
\centering
\includegraphics[scale=0.7]{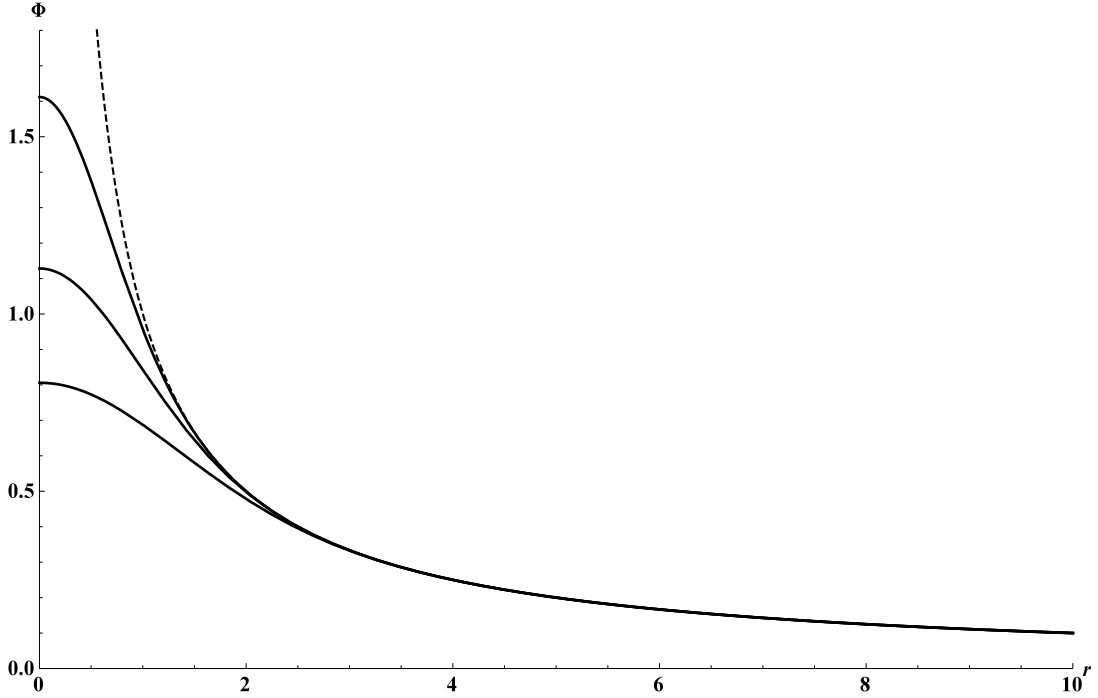}
\caption{Electrostatic potential generated by a point charge as a function of distance $r$, for some values of $\ell$. When $r \rightarrow 0$, the potential is finite at the origin.}
\label{GraficoPotencial}
\end{figure}

As we expect, the potential goes to zero when we consider large distances when compared to the $\ell$ scale, although in the limit $r \rightarrow 0$, the properties of the error function leads us to a finite potential at the origin

\begin{equation}\label{37}
\Phi_{0}=\frac{q}{4\pi \epsilon_{0}} \frac{1}{\ell \sqrt{2\pi}} \; .
\end{equation}
\ni Consequently, the electrostatic self-energy is also finite in this limit.  Concerning the electron case, where $q=-e$, we obtain the electron classic self-energy.
\begin{eqnarray} \label{38}
U_{e}=\frac{1}{2\sqrt{\pi}} \frac{e^2}{4\pi \epsilon_{0} \ell} \; .
\end{eqnarray}
Considering this minimum length scale, the electron can be seen as a sphere with radius equal to $\ell$, and total charge equal to $-e$.  It is well known in electrostatic, an homogeneous sphere, with radius $R$ and total charge $Q$ has the following energy
\begin{eqnarray}\label{39}
U=\frac{3}{5} \, \frac{Q^{2}}{4\pi \epsilon_{0} R} \; .
\end{eqnarray}
%

%
%
The final result of the electron energy differs from Eq. (\ref{39}) by a numerical factor $\sqrt{\pi}/2$, which shows that the electron can have a non-homogeneous charge distribution  which can be distributed through all its volume.  This fact indicates that the electron can be formed by sub-particles states, even more fundamental and non-uniformly distributed through all its volume.

The electric field concerning the point charge can be obtained with the Dirac delta function in Eq. (\ref{32})
\begin{eqnarray}\label{40}
\vec{E} (\vec{r}) = \frac{q}{4 \pi \epsilon_{0} \, r^2} \left[ \mbox{Erf} \left( \frac{r}{2\sqrt{2} \, \ell} \right) - \frac{r}{\ell \sqrt{2\pi}} \; e^{-{\left( \frac{r}{2 \sqrt{2} \, \ell} \right)^2}} \right] \hat{r} \; .
\end{eqnarray}
The flux of electric field through the Gaussian spherical surface $\partial {\cal R}$ of radius $r$, for a charge centered at the origin,
gives us the Gauss Law
\begin{eqnarray}
\oint_{\partial {\cal R}} \vec{E} \cdot \hat{\mathbf{n}} \, dA=\frac{1}{\epsilon_{0}} \, q_{eff}(r) \; ,
\end{eqnarray}
where we have defined the effective charge $q_{eff}$, as a function of the radial distance
\begin{eqnarray}
q_{eff}(r)=q \, \left[ \mbox{Erf} \left( \frac{r}{2 \sqrt{2} \, \ell} \right) - \frac{r}{\ell \sqrt{2\pi}} \; e^{-{\left( \frac{r}{2 \sqrt{2} \, \ell} \right)^2}} \right] \; .
\end{eqnarray}
If our point charge is interpreted as a sphere of radius $\ell$, the figure just below illustrates the distribution of charge as a function of a radial distance
$r$ throughout volume. This confirm the inhomogeneous distribution of charge discussed previously in the electron self-energy result.
\begin{figure}[h!]
\centering
\includegraphics[scale=0.6]{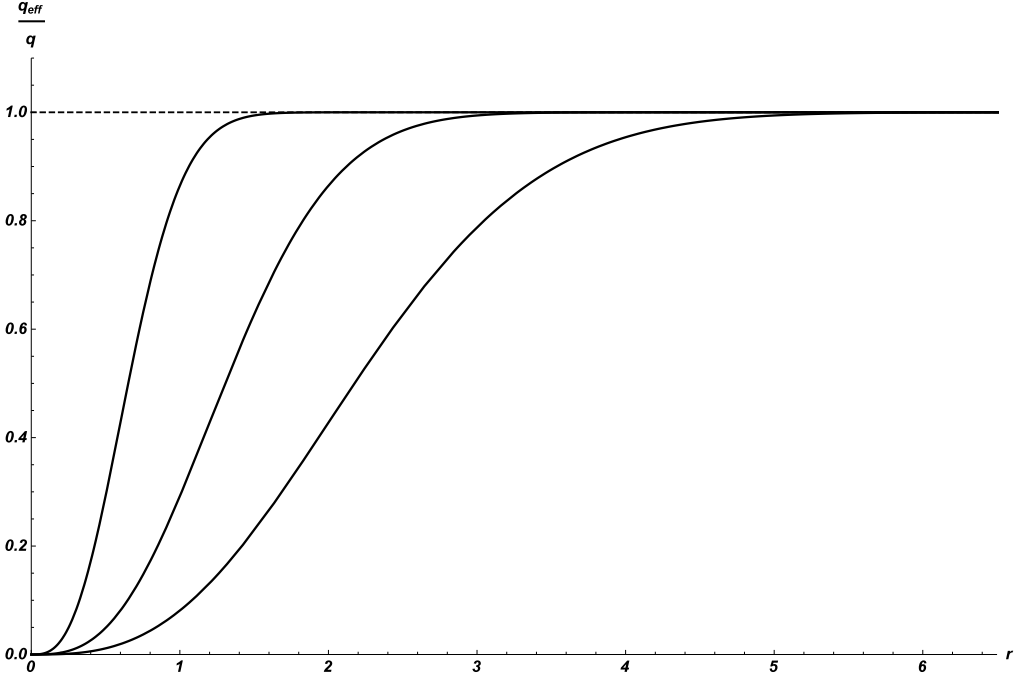}
\caption{ The  effective charge as function of radial distance $r$, for some values of $\ell$. When $r \rightarrow 0$, the effective charge goes to zero, and for $r \rightarrow \infty$, which obtain the asymptotic value $q_{eff}=q$.  }
\label{GraficoCarga}
\end{figure}
The electrostatic force between both charges separated by a distance $r$ is given by
\begin{eqnarray}\label{41}
\vec{F} (\vec{r}) = \frac{q^2}{4 \pi \epsilon_{0} \, r^2} \left[ \mbox{Erf} \left( \frac{r}{2 \sqrt{2} \, \ell} \right) - \frac{r}{\ell \sqrt{2\pi}} \; e^{-{\left( \frac{r}{2 \sqrt{2} \, \ell} \right)^2}} \right] \hat{r} \; .
\end{eqnarray}
Using the limit $\ell \rightarrow 0$, we can obtain the standard electrostatic field for a point charge.  And, in the force expression, the Coulomb law can also be recovered.  However, the interesting point is the limit $r \rightarrow 0$, where the expression which represents the force (\ref{41}) is zero at the origin.  This conclusion leads us to the analysis of the force in the region next the origin region.   To carry out an expansion in $r=0$ we have that
\begin{figure}[h]
\centering
\includegraphics[scale=0.7]{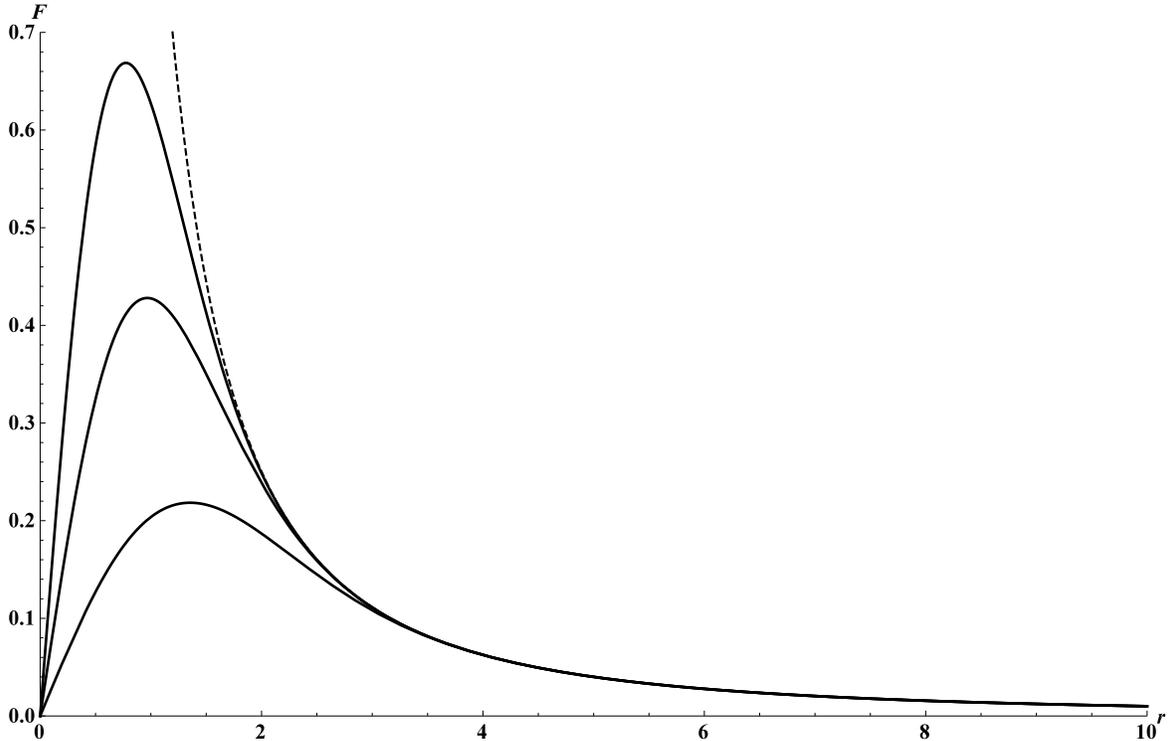}
\caption{The force between two charges as function of distance $r$ between both ones, for some values of $\ell$.  The dashed line shows the Colombian force.  In the region  $r \ll \ell$, the curve shows the difference between both forces.}
\label{GraficoForca}
\end{figure}
\begin{eqnarray}\label{42}
\vec{F}(r)=\frac{q^{2}}{4\pi \epsilon_{0}} \, \frac{\vec{r} }{6 \ell^{3} \sqrt{\pi}} \, \left( 1 - \frac{3}{20} \, \frac{r^{2}}{\ell^{2}} + \, \ldots \right) \; ,
\end{eqnarray}
\ni which shows a linear expression relating force and distance $r$.  Considering small distances, i.e., $r << 0$, the force is proportional to the distance $r$, namely,
$|\vec{F}(r)| \, \propto \, r$.  In this region for small distances, the force shows a confining feature.
%
%

\pagebreak

\section{The scalar field model}

In this section we will show the model of scalar field with self-interaction-$\phi^4$, some quantum properties of the model, and the influence of the minimal length on the its  mass. To do that, we will introduce the dispersion relation for a particle of mass $m$ associated with the representation (\ref{Pmunablamu}), namely, given by \footnote{From now on, we will adopt the natural units $c=\hbar=1$ by convenience.}
\begin{equation}\label{RelDisp}
p_{\mu}p^{\mu} \, e^{-2\ell^2 \, p_{\mu}p^{\mu}}=m^2 \; ,
\end{equation}
where $p^{\mu}=\left( \, E,\vec{p} \, \right)$ is the usual momentum in special relativity. Since the relativistic symmetry stays unaltered here,
the dispersion relation is invariant under Lorentz symmetry, the solution of (\ref{RelDisp}) is given by your inverse function
\begin{eqnarray}
p_{\mu}p^{\mu}=-\frac{1}{2\ell^2} \, W\left(-2 \, \ell^2 m^2 \right) \; .
\end{eqnarray}

Moreover, the propagation of fields with this new momentum representation must be affected, and obviously modifies the behavior of the
quantum field theory.   Hence, the next step is to construct the first field equations via the momentum operator representation (\ref{10}).
In this way, the Klein-Gordon equation concerning a scalar field $\varphi$ is given by
\begin{eqnarray}\label{50}
\left(\nabla_{\mu}\nabla^{\mu}+m^2\right)\varphi(x)=0 \; .
\end{eqnarray}
As a simple example, we can see the influence of the minimal length in the scalar model with quartic self-interaction $\phi^4$.
The Lagrangian of this model is given by
\begin{equation}\label{LPhi4}
{\cal L}_{Scalar-\varphi^4}=\frac{1}{2} \left(\nabla_{\mu}\varphi\right)^{2}-\frac{1}{2} \, m^{2} \, \varphi^2-\frac{g}{4!} \, \varphi^4 \; ,
\end{equation}
where $g$ is a real coupling constant. Integrating by parts we write it in the form field-operator-field
\begin{equation}\label{LPhi4}
{\cal L}_{Scalar-\varphi^4}=-\frac{1}{2} \, \varphi \left( \, e^{2\ell^2\Box} \, \Box + m^{2} \right) \, \varphi -\frac{g}{4!} \, \varphi^4 \; .
\end{equation}
It is easy to see that the Feynman's propagator in the momentum space has a divergence in the ultraviolet limit, when $p^2 \gg 1$.
%
%
%
%
However, it is convenient in this model to work in the Euclidean space-time, and making the usual
Wick's rotation, $p^2 \rightarrow -p_{E}^{2}$, so the propagator is given by
\begin{eqnarray}
\Delta_{F}(p_{E}^2)=\frac{1}{e^{2\ell^2 \, p_{E}^2} \, p_{E}^2 +m^{2}} \; ,
\end{eqnarray}
when $p_{E}^{\, 2} \gg 1$, this propagator falls down rapidly to zero and can be written as
\begin{eqnarray}
\Delta_{F}(p_{E}^{\, 2}) \sim \frac{e^{-2\ell^2 \, p_{E}^{\, 2}}}{p_{E}^{\, 2}} \; .
\end{eqnarray}
In scalar model $\varphi^4$, the first divergence diagram in the perturbative series is the tadpole contribution to
scalar propagator. By simplicity, we consider the massless scalar case represented by the integral
\begin{eqnarray}
\Sigma_{1}=\frac{ig}{2} \int \frac{d^4p_{E}}{\left(2\pi\right)^4} \; \frac{e^{-2\ell^2 \, p_{E}^2}}{p_{E}^2} \, .
\end{eqnarray}
By a simple power counting, this integral is well defined in the ultraviolet range, that is,
$\Sigma_{1} \sim p_{E}^{\, 2} \, e^{-2\ell^2 \, p_{E}^2}$, in which the exponential function makes a
natural cut-off. To make it explicitly, we use the Schwinger's integral
\begin{eqnarray}\label{IdSchwinger}
\frac{1}{p_{E}^{\, 2}}=\int_{0}^{\infty} ds \; e^{-s \, p_{E}^{\, 2}} \; ,
\end{eqnarray}
so the previous integral can be written as
\begin{eqnarray}
\Sigma_{1}=\frac{g}{2} \int_{0}^{\infty} ds \;
\int \frac{d^4p_{E}}{\left(2\pi\right)^4} \; e^{-\left(s +2\ell^2\right) \, p_{E}^2} \; .
\end{eqnarray}
Using the well known Gaussian integral, we obtain the finite result
\begin{eqnarray}
\Sigma_{1}= \frac{g}{4\left(16 \pi\right)^{2}} \, \frac{1}{\ell^2} \; .
\end{eqnarray}
Thus, the previous result is the finite contribution of perturbative series at the one loop approximation,
that can be interpreted as the ``mass" of scalar field, obtained by the introduction of the minimal length.
Explicitly, the propagator at the one loop is given by
\begin{eqnarray}
\Delta_{F}(p_{E}^{\, 2}) = \frac{e^{-2\ell^2 \, p_{E}^{\, 2}}}{p_{E}^{\, 2}+\frac{g}{4\left(16 \pi\right)^{2}} \, \frac{1}{\ell^2}} \; .
\end{eqnarray}
In this approximation the propagator has acquired a non trivial pole
\begin{eqnarray}
m_{\varphi}=\frac{\sqrt{g}}{32 \pi} \, \frac{1}{\ell} \; .
\end{eqnarray}
Thus, the mass of scalar is identified and it was regularized by the minimal length. This preliminary result
show us that new scale has worked like a natural regulator of the ultraviolet divergences. In the next section,
we will investigate an effective quantum electrodynamics, and the influence of $\ell$-scale in the electron mass.

%
%
\section{The effective quantum electrodynamics}

For a $1/2$-spin field $\psi$, the Dirac equation for the electron is given by
\begin{eqnarray}\label{51}
\left(i \gamma^{\mu}\nabla_{\mu} -m\right)\psi(x)=0 \; .
\end{eqnarray}
It is easy to see that the Lagrangian which lead us to the anterior equation is given by
\begin{eqnarray}\label{52}
{\cal L}_{Dirac}=\bar{\psi}(x)\left(i\gamma^{\mu}\nabla_{\mu}-m \right)\psi(x) \; .
\end{eqnarray}
\ni Our main motivation in this work is to construct an interaction model between leptons and the EM field $A^{\mu}$.  To accomplish that, we will use the equivalence concerning the covariant derivative
\begin{eqnarray}\label{53}
\partial_{\mu} \, \longmapsto D_{\mu}=\partial_{\mu} + ie \, Q  A_{\mu} \; ,
\end{eqnarray}
\ni where $Q$ is the charge generator associated with the symmetry group of the model, and the derivative $\nabla_{\mu}$ is substituted by the operator
\begin{eqnarray}\label{54}
\nabla^{(A)}_{\mu}:=e^{\ell^{2}\Box^{(A)}}D_{\mu} \; ,
\end{eqnarray}
\ni where $\Box^{(A)} := D_{\mu} D^{\mu}$. To obtain the gauge symmetry of the Lagrangian in Eq. (\ref{52}), we can provide the local transformation
$U(1)$
\begin{eqnarray}\label{55}
\psi \longmapsto \psi^{\prime}(x) &=& e^{ie \, Q \, \Lambda(x)} \; \psi(x) \; ,
\\
A_{\mu} \longmapsto A_{\mu}^{\; \prime} \! &=& \! A_{\mu}-\partial_{\mu}\Lambda(x) \; \nonumber ,
\end{eqnarray}
\ni where $\Lambda(x)$ is a real function of space-time. Using the set of transformations (\ref{55}),
the operator $\nabla_{A_{\mu}}$ transformation is
\begin{eqnarray}\label{57}
\nabla^{(A)}_{\mu} \longmapsto \nabla^{(A) \; \prime}_{\mu} = e^{ie \, Q \, \Lambda(x)} \, \nabla_{\mu}^{(A)} \, e^{-ie \, Q \, \Lambda(x)} \; .
\end{eqnarray}
\ni So, the gauge transformations define the unitary Abelian group $U(1)$ and the invariant Lagrangian under the $U(1)$
transformation group which is given by
\begin{eqnarray}\label{60}
{\cal L}_{QED}=-\frac{1}{4} \, F_{\mu\nu}F^{\mu\nu}+\bar{\psi}(x)\left(i\gamma^{\mu}\nabla_{\mu}^{\,(A)}-m \right)\psi(x) \; ,
\end{eqnarray}
\ni which is the QED Lagrangian in the presence of a minimal length scale $\ell$, and this result opens the path to quantize this model.
When $\ell \rightarrow 0$, the usual QED is immediately recovered. By convenience, concerning quantization, we will introduce the gauge fixing term
\begin{eqnarray}
{\cal L}_{gf}=-\frac{1}{2 \, \xi} \left(\nabla_{\mu}A^{\mu} \right)^2 \; ,
\end{eqnarray}
where $\xi$ is a real parameter. If we write in a free part plus the interaction of the fields,
firstly we obtain the electron propagator
\begin{eqnarray}
S_{F}(p_{E})= \frac{i}{\slash\!\!\!p_{E} \, e^{\ell^2 \; p_{E}^{\,2}}+m} \; ,
\end{eqnarray}
and posteriorly, the photon propagator is given by
\begin{eqnarray}
\Delta_{\mu\nu}(k_{E}^2)=\frac{e^{-2\ell^2 \, k_{E}^2}}{k_{E}^{\, 2}} \, \left[\eta_{\mu\nu}+\left(\xi-1 \right) \frac{k_{E\mu}k_{E\nu}}{k_{E}^{\, 2}} \right] \; .
\end{eqnarray}
The interaction sector is written in the form of series
\begin{eqnarray}
{\cal L}_{int}=-e \, \bar{\psi} \, \gamma_{\mu} A^{\mu} \, \psi-e \, \ell^{2}  \, \bar{\psi}\gamma^{\mu}A_{\mu}\Box\psi
-e \, \ell^{2} \, \bar{\psi}\gamma^{\mu}\Box A_{\mu}\psi+\ldots+{\cal O}(\ell^4) \; .
\end{eqnarray}
The first term is just the usual interaction, but the following terms are non-renormalizable interactions in the quantum approach
of the perturbative theory, where the coupling constants are extremely weak, of orders $e \ell^2$, $e\ell^4$ and so on. Therefore,
we will just consider the usual interaction of the QED, in which we have an effective model where the minimal length emerges only in the
propagators expressions. Furthermore, we also consider the massless QED, where we had put $m=0$ in the electron Lagrangian. It may be interesting to see
the contribution of the minimal length to electron mass via electron self-energy. Using all rules in the Feynman gauge $(\xi=1)$,
the electron self-energy at the one loop is represented by the integral
\begin{equation}\label{Sigmaeletron}
-i\Sigma(\slash\!\!\!p_{E})=i(ie)^{2}\int\frac{d^{4}k_{E}}{(2\pi)^{4}} \, \frac{e^{-2\ell^2 \, k_{E}^2}}{k_{E}^{\, 2}}
\, e^{-\ell^2 \, \left(k_{E}-p_{E}\right)^2} \, \frac{\gamma_{\mu}\left(\slash \!\!\! p_{E}-\slash \!\!\! k_{E}\right)\gamma^{\mu}}{\left(k_{E}-p_{E}\right)^{2}} \; .
\end{equation}
By simple power counting, this integral is well defined in the ultraviolet regime due to exponential decay function.
In contrast, the infrared (IR) divergence will emerge here, at $k_{E}^{2}=0$.
This integration also suggests us to introduce the Schwinger identity (\ref{IdSchwinger}), and we can write it in the form
\begin{eqnarray}\label{Sigmaeletronintsdk}
-i\Sigma(\slash\!\!\!p_{E}) \! & \stackrel{\Lambda^{2} \rightarrow 0}{=} & \! -2i \, (ie)^{2} \int_{0}^{1/\Lambda^{2}} ds_{1} \int_{0}^{\infty} ds_{2} \int \frac{d^{4}k_{E}}{(2\pi)^{4}}
\, \times
\nonumber \\
&& \times \,
\left(\slash \!\!\! p_{E}-\slash \!\!\! k_{E}\right) \, e^{-\left(3\ell^2+s_{1}+s_{2}\right) \, k_{E}^2+2\, \left(\ell^2+s_{2}\right) \, k_{E}\cdot p_{E} -\left( \ell^2+s_{2} \right)p_{E}^{\, 2}} \; ,
\end{eqnarray}
where $\Lambda^2$ a is cut-off regulator with energy dimension to the square to control the IR divergence.
After a Gaussian integral, we can obtain that
\begin{eqnarray}\label{Sigmaeletronints12}
\Sigma(\slash\!\!\!p_{E}) \stackrel{\Lambda^{2} \rightarrow 0}{=} -2 \, \frac{e^2}{(4\pi)^{2}} \, \slash \!\!\!p_{E} \int_{0}^{1/\Lambda^{2}} ds_{1} \left( 2\ell^2+s_{1} \right)
\int_{0}^{\infty} ds_{2} \; \; \frac{e^{-\frac{(2\ell^2+s_1)(\ell^2+s_2)}{3\ell^2+s_1+s_2} \, p_{E}^{\, 2}}}{\left(3\ell^2+s_{1}+s_{2}\right)^{3}}
\; .
\end{eqnarray}
Since we have considered the massless electron, the on-shell condition $p_{E}^{\; 2}=0$ must be implemented here, and the exponential in the (\ref{Sigmaeletronints12}) is simplified to given the result
\begin{eqnarray}\label{SigmaeletronResult}
\Sigma(\slash\!\!\!p_{E}) \stackrel{\Lambda^{2} \rightarrow 0}{=} \, \frac{e^2}{3(4\pi)^{2}} \, \slash \!\!\!p_{E} \,
+ \, \frac{e^2}{(4\pi)^{2}} \, \slash \!\!\!p_{E} \, \ln\left(3\ell^{2}\Lambda^2\right) \; .
\end{eqnarray}
We have the finite part of electron self-energy plus the IR divergence term. It is important here that the minimal
length scale does not appear at the one loop approximation, as well as we have not a divergence of ultraviolet nature.
This IR divergence must be canceled out by the vertex correction at the one loop.


\section{Conclusions and perspectives}

One of the greatest challenges in theoretical physics today is to understand how the gravitational effects can be affected by quantum phenomena.  Namely, how can gravity and quantum mechanics interact in order to describe certain phenomena that occurred in the early Universe, for example.

The introduction of a minimal length  is one attempt to fathom such problem.  To describe our Universe through a discrete space-time instead of a continuous one has ignited an intense investigation of the theories described in this so-called NC space-time, which has multiple ways to depict noncommutativity like the canonical or Euclidean ways, or Snyder and kappa-deformed manners, to cite a few.

In this work we have introduced a minimal length, a quantity having length dimension which value is measured at the Planck scale.  One way to introduce the minimal length is through the so-called generalized uncertainty principle.  In this work we have deformed the coordinates and momentum commutation relations using the Quesne-Tkachuk algebra.  The commutative commutation relations can be recovered when the main parameters are put to be equal to zero, which is equivalent to put the NC parameter to be equal to zero in order to obtain also the standard commutative relations for the recovered commutative coordinates.

Since one of the motivations to introduce a minimal length is to regularize field theories, in this work we have analyzed the regularization of the standard electrodynamics through the construction of a electrodynamics with a minimal length.
Besides this not so original objective, the original thing here is that we have introduced a new way to modify electrodynamics through this minimal length.  The first objective would be to study the electrodynamics at Planck scale, as we explained just above.  In order to obtain the analytical results using this new way to introduce this minimal length we have solved the Green functions deformed by this minimal length.  Consequently, the regularization of the theory was obtained.  The mass of the scalar field was regularized and an interesting result was obtained.

After that, we have constructed the electron and the photon propagators as functions of the minimal length.  Our main result, the calculation of the electron self-energy at one loop as a function of the minimal length, has showed that we have the finite part of the electron self-energy plus the IR divergence term. We have seen that the minimal length did not appear at one loop approximation, and the result did not show a divergence term due to the UV nature of the theory.  We have shown that the IR divergence can be canceled out by the vertex correction at one loop.

As future perspective we can apply the new method to introduce a minimal length in theories with high derivative orders, as well into NC theories.  These perspectives are in fact ongoing research topics which will be published elsewhere.

\section{Acknowledgments}

\ni E.M.C.A. thanks CNPq (Conselho Nacional de Desenvolvimento Cient\' ifico e Tecnol\'ogico), Brazilian scientific support federal agency, for partial financial support through Grants No. 301030/2012-0 and No. 442369/2014-0 and the hospitality of Theoretical Physics Department at Federal University of Rio de Janeiro (UFRJ), where part of this work was carried out.

\newpage


\begin{thebibliography}{99}


\bibitem{string}   G. Veneziano, Europhys. Lett. 2 (1986) 199;
D. Amati, M. Ciafaloni and G. Veneziano, Phys. Lett. B 216 (1989) 41;
 Phys. Lett. B 197 (1987) 81;
 Int. J. Mod. Phys. A 7 (1988) 1615;
 Nucl. Phys. B bf 347 (1990) 530;
D.J. Gross and P.F. Mende, Nucl. Phys. B 303 (1988) 407;
K. Konishi, G. Paffuti and P. Provero, Phys. Lett. B 234 (1990) 276.

\bibitem{cls}    S. Capozziello, G. Lambiase and G. Scarpetta, Int. J. Theor. Phys. 39 (2000) 15.

\bibitem{garay}    L. J. Garay, Int. J. Mod. Phys. A 10 (1995) 145.

\bibitem{Heisenberg38} W. Heisenberg, Ann. Physik 32 (1938) 20.

\bibitem{Snyder47} H. S. Snyder, Phys. Rev. 71 (1947) 38.

\bibitem{yang47} C. N. Yang, Phys. Rev. 72 (1947) 874.

\bibitem{seibergwitten99} N. Seiberg and E. Witten, JHEP 9909 (1999) 032.

\bibitem{Szabo03} R. Szabo, Phys. Rep. 378 (2003) 207.

\bibitem{Moyaedi2013} S. Moyaedi, M. R. Setare and B. Khosropour, Adv. High En. Phys. (2013) ID 657870.

\bibitem{Quesne2006JPA} C. Quesne and V. M. Tkachuk, Journal of Physics A 39 (2006) 10909-10922.

\bibitem{Quesne2006Csz} C. Quesne and V. M. Tkachuk, Czechoslovak Journal of Physics 56 (2006) 1269-1274.


\bibitem{Kempf94} A. Kempf, Journal of Mathematical Physics 35 (1994) 4483.

\bibitem{Kempf95} A. Kempf, G. Mangano and R. B. Mann, Physical Review D 52 (1995) 1108.

\bibitem{Kempf96}  A. Kempf and H. Hinrichsen, Journal of Mathematical Physics 37 (1996) 2121.

\bibitem{Kempf97} A. Kempf, J. Phys. A: Math. Gen. 30 (1997) 2093–2101.

\bibitem{Chang2002} L. N. Chang, D. Minic, N. Okamura and T. Takeuchi, Physical Review D 65 (2002) 125027 1-8.

\bibitem{LeeWick69} T. D. Lee and G. C. Wick, Nuclear Physics B 9 (1969) 209-243.

\bibitem{NevesTurcati2014} R. Turcati and M.J. Neves, Advances in High Energy Physics (2014) ID 53953.

\bibitem{Accioly2011} A. Accioly, P. Gaete, J. A. Hela\"{y}el-Neto, E. SCATENA and R. Turcati,
Mod. Phys. Lett. A 26 (2011) 1985-1994.

\bibitem{Gradshteyn}      I. S. Gradshteyn, I. M. Ryzhik, {\it Table of integrals, series and products}, seventh edition 2007.

\bibitem{nekrasov}   M. R. Douglas and N. A. Nekrasov, Rev. Mod. Phys. 73 (2001) 977.

\bibitem{szabo}   R. Szabo, Gen. Rel. Grav. 42 (2010) 1; Class. Quant. Grav. 23 (2006) R199.

\bibitem{abreuneves}  M. J. Neves and E. M. C. Abreu, Nucl. Phys. B884 (2014) 741;
Int. J. Mod. Phys. A 28 (2013) 1350017;
Int. J. Mod. Phys. A 27 (2012) 1250109; ``The Standard Electroweak Model in the Noncommutative DFR Space-Time,"  arXiv:1511.06450; \\
``Spontaneous symmetry breaking and masses numerical results in DFR noncommutative space-time," arXiv:1510.08354; \\
``The Yang-Mills gauge theory in DFR noncommutative space-time," arXiv:1506.00035; \\
E. M. C. Abreu, M. J. Neves and V. Nikoofard, ``New remarks on DFR noncommutative phase-space," arXiv:1501.01912, and references within.

\end{thebibliography}
%
%

\end{document}